\documentclass[preprint]{aastex}

\slugcomment{Version 4 June 2002}

\newcommand\al{\alpha}

\renewcommand\th{\theta}

\newcommand\phir{\phi_{r}}
\newcommand\phix{\phi_{x}}
\newcommand\phiy{\phi_{y}}
\newcommand\phiij{\phi_{ij}}
\newcommand\phixx{\phi_{xx}}
\newcommand\phiyy{\phi_{yy}}
\newcommand\phixy{\phi_{xy}}
\newcommand\phirr{\phi_{rr}}

\newcommand\Sigcr{\Sigma_{\rm cr}}
\newcommand\gravlens{{\it gravlens\/}}

\newcommand\xx{{\bf x}}

\newcommand\yy{{\bf y}}
\newcommand\uu{{\bf u}}
\newcommand\bb{{\bf b}}

\newcommand\del{{\bf\nabla}}

\newcommand\refeq[1]{eq.~(\ref{eq:#1})}
\newcommand\refEq[1]{Eq.~(\ref{eq:#1})}
\newcommand\refeqs[2]{eqs.~(\ref{eq:#1}) and (\ref{eq:#2})}
\newcommand\refEqs[2]{Eqs.~(\ref{eq:#1}) and (\ref{eq:#2})}

\newcommand\reftab[1]{Table~\ref{tab:#1}}

\begin{document}

\title{A Catalog of Mass Models for Gravitational Lensing}
\author{Charles R.\ Keeton\footnote{Hubble Fellow}}
\affil{Astronomy and Astrophysics Department, University of Chicago, \\
5640 S.\ Ellis Ave., Chicago, IL 60637}

\begin{abstract}
Many different families of mass models are used in modern
applications of strong gravitational lensing.  I review a
wide range of popular models, with two points of emphasis:
(1) a discussion of strategies for building models suited
to a particular lensing problem; and (2) a summary of
technical results for a canonical set of models.  All of
the models reviewed here are included in publicly-available
lensing software called \gravlens.
\end{abstract}

\section{Introduction}

As applications of strong gravitational lensing have become more
sophisticated, the variety and complexity of mass distributions
used for lensing studies have increased.  Gone are the days when
the singular isothermal sphere was all you needed to know.  Now
we have softened power law ellipsoids, pseudo-Jaffe models,
generalized NFW models, exponential disks embedded in isothermal
halos, and so on.  Plus, all of these models are likely to be
perturbed by other galaxies, groups, or clusters near the lens
galaxy or along the line of sight.

I have developed software called \gravlens\ for a variety of
lensing and lens modeling applications.\footnote{The software
is discussed in a separate paper (Keeton 2001), and is available
to the community via the web site of the CfA/Arizona Space
Telescope Lens Survey, at {\tt http://cfa-www.harvard.edu/castles}.}
In the course of writing the code I have collected most of the
mass models used for lensing studies, and this paper reviews
those models.  The outline is as follows.  Section 2 discusses
circular and elliptical symmetry in lensing mass distributions,
and argues that a canonical set of circular and elliptical models
provides a useful basis set for building much more complex
composite models.  Section 3 gives some suggestions (propaganda,
really) about selecting models appropriate for your application.
Section 4 presents the general equations that describe the lensing
properties of a given mass distribution.  Finally, Section 5
presents a detailed catalog of results for a number of models.
The mass models discussed in this paper (and included in the
\gravlens\ software) are listed in \reftab{mods}.

\section{Circular, Elliptical, and Composite Lens Models}

The lensing properties of any mass distribution can be written
in terms of two-dimensional integrals over the surface mass
density (see \S 4).  In general the integrals cannot be
evaluated analytically, but many lensing applications offer
simplifications due to symmetry.  Applications such as
microlensing may permit the use of a lens with circular
symmetry, in which case the lensing properties can usually be
found analytically.  In other applications, such as lensing by
galaxies, it may be reasonable to assume elliptical symmetry,
which allows the lensing properties to be written as a set of
one-dimensional integrals (see \S 4); the integrals can
sometimes be evaluated analytically and are always amenable to
fast numerical techniques.

The geometric simplifications may not hold in increasingly
sophisticated lensing applications. For example, companion stars or
planets can break circular symmetry in microlensing (e.g., Mao \&
Paczy\'nski 1991; Gould \& Loeb 1992); neighboring galaxies (e.g.,
Young et al.\ 1981; Schechter et al.\ 1997; Koopmans \& Fassnacht
1999) or the lens galaxy's internal structure (e.g., Maller,
Flores \& Primack 1997; Bernstein \& Fischer 1999; Keeton et al.\
2000) may break elliptical symmetry in lensing by galaxies; and
individual cluster members can ruin the symmetry of the smooth
background potential in lensing by clusters (e.g., Bartelmann \&
Steinmetz 1995; Tyson, Kochanski \& Dell'Antonio 1998).
Nevertheless, in all of these examples the total mass distribution
can be written as a combination of circular or elliptical components,
either placed at different positions to represent different objects,
or combined at the same position to mimic an object with more
complicated internal structure (also see Schneider \& Weiss 1991).
Because the Poisson equation is linear, the lensing potential of
such composite models is simply the sum of the component potentials.
In other words, arbitrary combinations of circular and elliptical
models yield a wide range of complex mass distributions whose
lensing properties are nevertheless easy to compute.

While composite models provide a great deal of freedom and
complexity, they are not completely general. This limitation is
eliminated in the elegant algorithm by Saha \& Williams (1997; also
Williams \& Saha 2000) for finding non-parametric lens models. The
approach is to introduce a set of mass pixels and construct a
large linear programming problem for determining their masses.
The problem is severely underconstrained, but by requiring
positive-definite masses and imposing some smoothness criteria
it is possible to find a wide but finite range of non-parametric
models consistent with the data.

A limitation of the Saha \& Williams algorithm is that the
constraint of a positive-definite surface density is weaker than
the constraint of a positive 3-d mass density --- or, better yet,
a positive-definite quasi-equilibrium distribution function. In
other words, while parametric models may provide too little
freedom, the Saha \& Williams non-parametric models provide too
much. For example, their method does not determine whether models
are consistent with stellar dynamics, and many of the models found
by the method probably are not consistent. The \gravlens\ software
includes only parametric models that have physical motivations
apart from lensing, but it offers the ability to combine them
in arbitrary ways to achieve extensive and reasonable complexity.

\section{Selecting Models}

Selecting the class of models to use for a particular application
is a key part of lens modeling. When modeling lenses produced by
galaxies, a simple and useful place to start is an isothermal
model, i.e.\ a model with density $\rho \propto r^{-2}$ and a
flat rotation curve. Spiral galaxy rotation curves (e.g., Rubin,
Ford \& Thonnard 1978, 1980), stellar dynamics of elliptical
galaxies (e.g., Rix et al.\ 1997), X-ray halos of elliptical
galaxies (Fabbiano 1989), models of some individual lenses (e.g.,
Kochanek 1995; Cohn et al.\ 2001), joint lensing and dynamical
analyses (e.g., Treu \& Koopmans 2002; Koopmans \& Treu 2002),
and lens statistics (e.g., Maoz \& Rix 1993; Kochanek 1993, 1996)
are all consistent with roughly isothermal profiles.

However, an isolated isothermal ellipsoid
rarely yields a good quantitative fit to observed lenses (e.g.,
Keeton et al.\ 1997, 1998; Witt \& Mao 1997). In general, adding
parameters to the radial profile of the galaxy fails to produce a
good fit, but adding parameters to the angular structure of the
potential dramatically improves the fit (e.g., Keeton \& Kochanek
1997; Keeton et al.\ 1997). The additional angular structure comes
from the tidal perturbations of objects near the main lens galaxy
or along the line of sight. In other words, the fact that few
galaxies are truly isolated means that lens models generically
require two independent sources of angular structure: an
ellipsoidal galaxy plus external perturbations. The combination of
angular terms can make it difficult to disentangle the shape of the
galaxy and the nature of the external perturbations, and it is
extremely important to understand any degeneracies between the two
sources of angular structure before drawing conclusions from the
models (see Keeton et al.\ 1997).

To move beyond isothermal models and explore other radial
profiles, softened power law lens models have traditionally been
very popular. However, these models have flat cores, while many
early-type galaxies have cuspy luminosity distributions (e.g.,
Faber et al.\ 1997), and dark matter halos in cosmological
simulations have cuspy mass distributions (e.g., Navarro, Frenk
\& White 1996, 1997; Moore et al.\ 1998, 1999). The lack of
central or ``odd'' images in most observed galaxy lenses also
limits the extent to which galaxies can have flat cores (e.g.,
Wallington \& Narayan 1993; Rusin \& Ma 2001). Cohn et al.\
(2001) thus argue that softened power law models are outdated
and should be replaced with families of cuspy lens models.
There are three traditional families of models in which the
profile of the cusp is fixed: NFW (Navarro, Frenk \& White 1996,
1997), Hernquist (1990), and de Vaucouleurs (1948) models. In
addition, there are three families of models in which the cusp
is taken to be an arbitrary power law: Nuker law models (Lauer
et al.\ 1995; Byun 1996; Keeton 2002), and two variants of NFW
models (see \S 5 for definitions; Jing \& Suto 2000; Keeton \&
Madau 2001; Mu\~noz, Kochanek \& Keeton 2001; Wyithe, Turner \&
Spergel 2001).

An important reason to study families of cuspy lens models is to
test the prediction from the Cold Dark Matter paradigm that halos
at a wide range of masses are consistent with a unified family
of halo models (e.g., Navarro et al.\ 1996, 1997).  The
distribution of image separations among the known lenses rules
out the hypothesis that galaxies and clusters can be described
by identical lens models (Keeton 1998; Porciani \& Madau 2000;
Kochanek \& White 2001).  However, it may be possible to resolve
the conflict with a model in which galaxies and clusters start
out with similar halo profiles, but baryonic processes such as
cooling modify the inner profiles of galaxies.  Kochanek \&
White (2001) show that such a model can match the observed image
separation distribution, but it remains to be seen whether the
model agrees with the lack of central or ``odd'' images in
observed lenses, or with detailed models of individual lenses.

\section{General Equations}

For a mass distribution with surface mass density $\kappa(\xx) =
\Sigma(\xx)/\Sigcr$ in units of the critical surface density for
lensing, the two-dimensional lensing potential is (e.g., Schneider,
Ehlers \& Falco 1992)
\begin{equation}
  \phi(\xx) = {1 \over \pi} \int \ln|\xx-\yy|\ \kappa(\yy)\ d\yy\,.
\end{equation}
The other lensing properties can be derived from the potential.
The deflection angle $\del\phi$ determines the positions of
images via the lens equation,
\begin{equation}
  \uu = \xx - \del\phi(\xx)\,,
\end{equation}
where $\uu$ is the source position.  The magnification tensor,
\begin{equation}
  \mu \equiv \left({\partial\uu \over \partial\xx}\right)^{-1}
  = \left[\begin{array}{cc}
    1 - \phixx & -\phixy \\
    -\phixy & 1 - \phiyy \\
  \end{array}\right]^{-1} ,
\end{equation}
determines the distortions and brightnesses of images.  
(Subscripts denote partial differentiation, $\phiij \equiv
\partial^2 \phi / \partial x_i \partial x_j$.)  Many lensing
applications involve only the locations and brightnesses of the
images, and thus require only the deflection and magnification
components. Applications that involve the time delays, such as
lensing measurements of the Hubble constant $H_0$, also require
the potential.

If the mass distribution has circular symmetry, the deflection
vector is purely radial and has an amplitude given by the 1-d
integral
\begin{equation} \label{eq:circdef}
  \phir(r) = {2 \over r} \int_0^r u\,\kappa(u)\,du
  = {1 \over \pi \Sigcr}\,{M_{\rm cyl}(r) \over r}\,,
\end{equation}
where $M_{\rm cyl}(r)$ is the mass enclosed by a cylinder of radius
$r$ (the projected mass), which is often easily evaluated. The
potential and magnification components can be obtained by
integrating or differentiating $\phir$.

More general is the case of elliptical symmetry, in which the
surface mass density has the form
\begin{equation} \label{eq:xi}
  \kappa = \kappa(\xi), \quad
  \mbox{where}\quad \xi^2 = x^2+y^2/q^2\,,
\end{equation}
where $q$ is the projected axis ratio and $\xi$ is an ellipse
coordinate. This is the functional form in a coordinate system with
the ellipse centered on the origin and aligned along the $x$-axis;
other coordinate systems can be reached by suitable translation and
rotation.  With elliptical symmetry the lensing properties can
be written as a set of 1-d integrals (see Schramm 1990, although I
have changed variables in the integrals),
\begin{eqnarray}
  \phi  (x,y) &=& {q\over2}\,I(x,y)                  \label{eq:phi1d}   \\
  \phix (x,y) &=& q\,x\,J_0(x,y)                     \label{eq:phix1d}  \\
  \phiy (x,y) &=& q\,y\,J_1(x,y)                     \label{eq:phiy1d}  \\
  \phixx(x,y) &=& 2\,q\,x^2\, K_0(x,y) + q\,J_0(x,y) \label{eq:phixx1d} \\
  \phiyy(x,y) &=& 2\,q\,y^2\, K_2(x,y) + q\,J_1(x,y) \label{eq:phiyy1d} \\
  \phixy(x,y) &=& 2\,q\,x\,y\,K_1(x,y)               \label{eq:phixy1d}
\end{eqnarray}
where the integrals are
\begin{eqnarray}
  I  (x,y) &=& \int_0^1 {\xi(u) \over u}\,{\phir\left(\xi(u)\right) \over
               \left[1-(1-q^2)u\right]^{1/2}}\,du \label{eq:ipot} \\
  J_n(x,y) &=& \int_0^1 {\kappa\left(\xi(u)^2\right) \over
               \left[1-(1-q^2)u\right]^{n+1/2}}\,du \\
  K_n(x,y) &=& \int_0^1 {u\,\kappa'\!\left(\xi(u)^2\right) \over
               \left[1-(1-q^2)u\right]^{n+1/2}}\,du \\
\mbox{where}\quad
  \xi(u)^2 &=& u\,\left( x^2 + {y^2 \over 1-(1-q^2) u } \right)
\end{eqnarray}
and $\kappa'(\xi^2) = d\kappa(\xi^2)/d(\xi^2)$. Note from
\refeq{ipot} that the potential can be written as an integral over
the circular deflection function $\phir$ from \refeq{circdef}, but
$\phir$ must be evaluated at the appropriate ellipse coordinate
$\xi(u)$.

All of the previous expressions assume that the surface density
$\kappa$ is known. Some models have 3-d density distributions for
which the projection integral cannot be evaluated analytically.
In this case even a spherical lens model requires computationally
expensive double integrals (the projection integral followed by
the lensing integral). However, the double integrals can be
rewritten as follows so the projection integral is replaced by
the enclosed mass $M(r)$, which can often be computed analytically.
(The mass $M(r)$ is the mass in spheres, which is different from
$M_{\rm cyl}(r)$ in eq.~\ref{eq:circdef}.) Writing $\kappa(r)$ as
an integral over $\rho(r)$ --- the projection integral --- and
substituting $\rho(r) = M'(r)/(4\pi r^2)$ where $M'(r) = dM/dr$,
we find
\begin{eqnarray}
  \kappa(r)
  &=& {1 \over 2 \pi \Sigcr} \int_r^\infty du\, {M'(u)
    \over u \sqrt{u^2-r^2}}\ , \label{eq:kap1} \\
  &=& {1 \over 2 \pi \Sigcr r} \int_0^1 dy\, {1 \over 1+y^2}\,
    \left[ M'\left(r\sqrt{1+y^2}\right) +
    M'\left(r\sqrt{1+y^{-2}}\right) \right], \label{eq:kap2}
\end{eqnarray}
where the second line represents a change of variables so the
integral has a finite range, which is convenient for numerical
integration. Combining \refeqs{circdef}{kap1} gives the circular
deflection as
\begin{eqnarray}
  \phir &=& {1 \over \pi \Sigcr r} \int_0^r du\,u
    \int_u^\infty dv\,{M'(v) \over v \sqrt{v^2-u^2}}\ , \\
  &=& - {1 \over \pi \Sigcr r} \int_0^r du\,u
    {\partial \over \partial u} \int_u^\infty dv\,M'(v)\,
    {\sqrt{v^2-u^2} \over v}\ , \\
  &=& {1 \over \pi \Sigcr r}\,\left[ M_\infty -
    \int_r^\infty dv\,M'(v)\, {\sqrt{v^2-u^2} \over v} \right] ,
\end{eqnarray}
where $M_\infty$ is the total mass. Integrating by parts then
yields
\begin{eqnarray}
  \phir &=& {r \over \pi \Sigcr} \int_r^\infty dv\,
    {M(v) \over v^2 \sqrt{v^2-r^2}}\ , \label{eq:def1} \\
  &=& {1 \over \pi \Sigcr r} \int_0^1 dy\,
    {1 \over (1+y^2)^{3/2}} \left[ M\left(r\sqrt{1+y^2}\right) +
    y\,M\left(r\sqrt{1+y^{-2}}\right) \right], \label{eq:def2}
\end{eqnarray}
where again the second line represents a change of variables for
numerical integration. \refEqs{kap1}{def1}, or \refeqs{kap2}{def2}
for numerical integration, represent the desired 1-d integrals for
the surface density and deflection. The magnification also requires
$\phirr$, which could be computed by differentiating $\phir$;
however, it is easier to compute $\kappa$ and $\phir$ and then use
the identity $r^{-1}\,\phir + \phirr = 2 \kappa$ to determine
$\phirr$.

\section{The Catalog}

\reftab{mods} lists a wide variety of popular lens models, all
of which are available in the \gravlens\ software.  This section
summarizes what is known about the mass distributions and lensing
properties of the models.  Analytic results are given where
available, which includes all but one of the circular models and
some of the elliptical models. Note that if the potential $\phi$ is
regular at the origin, it is normalized to have $\phi(0)=0$.
Lensing is insensitive to an arbitrary constant added to the
potential.

I have tried to include relevant references.  {\it If you use
results given here, please cite the original references rather
than this catalog.\/}  As for the unreferenced results, some
of them are new, while others are derived easily enough that
references seem unnecessary.  Use your own judgement about
citing such material.

\def\crc{\mbox{circular:}\quad}
\def\ell{\mbox{elliptical:}\quad}

{\bf Point mass:}
This model is inherently circular. A point mass $M$ produces a
lensing potential
\begin{eqnarray}
\phi = R_E^2\,\ln r\,,
\end{eqnarray}
where the Einstein radius is (in angular units)
\begin{equation}
  R_E = \sqrt{ {4 G M \over c^2}\,{D_{\rm ls} \over D_{\rm ol} D_{\rm os}} }\ .
\end{equation}

{\bf Softened power law potential:}
Sometimes for analytic simplicity it is convenient to put the
elliptical symmetry in the potential rather than in the density
(e.g., Blandford \& Kochanek 1987; Witt \& Mao 1997, 2000; Evans
\& Hunter 2002),
\begin{equation}
  \phi = b\,\left(s^2+x^2+y^2/q^2\right)^{\al/2} - b\,s^{\al}\, .
\end{equation}
The deflection and magnification are given by simple derivatives.

{\bf Softened power law ellipsoid:}
This model has a projected surface density
\begin{equation}
  \kappa(\xi) = {1 \over 2}\,{b^{ 2-\al} \over
    \left(s^2+\xi^2\right)^{1-\al/2} }\ ,
\end{equation}
which represents a flat core with scale radius $s$, and then a
power law decline with exponent $\al$ defined such that the mass
grows as $M_{\rm cyl}(r) \propto r^\al$ asymptotically. The core
radius can be zero if $\al>0$. The model gives a softened isothermal
model for $\al=1$, a modified Hubble model for $\al=0$, and a
Plummer model for $\al=-2$ (see Binney \& Tremaine 1987).  The
circular model has lensing properties
\begin{eqnarray}
\phi  &=& {1\over\al^2}\,b^{2-\al}\,r^\al\, {}_2 F_1\left[-{\al\over2} ,
          -{\al\over2} ; 1-{\al\over2} ; -{s^2\over r^2}\right]
          - {1\over\al}\,b^{2-\al}\,s^\al\,\ln\left({r \over s}\right) \nonumber \\
       &&\quad - {1\over2\al}\,b^{2-\al}\,s^\al\,\left[ \gamma_E -
            \Psi\left(-{\al\over2}\right) \right] \label{eq:alphapot} \\
      &=& {1\over\al^2}\,b^{2-\al}\,r^\al \qquad (\al>0, s=0) \\
\phir &=& {b^{2-\al} \over \al r}\,\left[ \left(s^2+r^2\right)^{\al/2}
  -s^\al \right] \qquad (\al \ne 0) \\
      &=& {b^2 \over r} \ln\left( 1+{r^2 \over s^2} \right) \qquad (\al = 0)
\end{eqnarray}
In the potential, ${}_2 F_1[a,b;c;x]$ is a hypergeometric function,
which can be written as or transformed into a quickly converging
series (see Press et al.\ 1992; Gradshteyn \& Ryzhik 1994, \S 9.1).
Also, $\gamma_E = 0.577216\ldots$ is Euler's constant, and $\Psi(x)
= d[\ln\Gamma(x)]/dx$ is the digamma function, or the logarithmic
derivative of the factorial function $\Gamma(x)$. Analytic
solutions for the elliptical model are possible for $\al = 0,
\pm1$, and two of these are given below. Barkana (1998) gives a
fast numerical algorithm for general softened power law ellipsoid
models.

{\bf Isothermal ellipsoid, $\al=1$:}
This model describes mass distributions with flat rotation curves
(outside the core). Its lensing properties are:
\begin{eqnarray}
\crc
\phi  &=& r\,\phir - b\,s\,\ln\left({ s+\sqrt{s^2+r^2}
          \over 2s }\right) \\
\phir &=& {b \over r}\,\left(\sqrt{s^2+r^2}-s\right) \\
\ell
\phi  &=& x\,\phix + y\,\phiy - b\,q\,s\,\ln\left[
              (\psi+s)^2 + (1-q^2) x^2 \right]^{1/2} \nonumber \\
       &&\quad +\,b\,q\,s\,\ln\left[(1+q)s\right]
      \label{eq:isophi} \\
\phix &=& {b\,q \over \sqrt{1-q^2}}\,\mbox{tan}^{-1}\left[
              { \sqrt{1-q^2}\,x \over \psi + s } \right]
      \label{eq:isophix} \\
\phiy &=& {b\,q \over \sqrt{1-q^2}}\,\mbox{tanh}^{-1}\left[
              { \sqrt{1-q^2}\,y \over \psi + q^2 s } \right]
      \label{eq:isophiy}
\end{eqnarray}
where $\psi^2 = q^2(s^2+x^2)+y^2$. The elliptical solutions have
been given by Kassiola \& Kovner (1993), Kormann, Schneider \&
Bartelmann (1994), and in the simple form quoted here by Keeton \&
Kochanek (1998). In the limit of a singular ($s=0$) and spherical
($q=1$) model, $b$ is the Einstein radius of the model and is
related to the 1-d velocity dispersion $\sigma$ by
\begin{equation}
  b = 4\pi \left({\sigma \over c}\right)^2\,{D_{\rm ls} \over D_{\rm os}}
\end{equation}
(in angular units).

{\bf $\al=-1$ ellipsoid:}
This model corresponds to an unnamed density profile with $\Sigma
\propto r^{-3}$ ($\rho \propto r^{-4}$) asymptotically. Its lensing
properties are:
\begin{eqnarray}
\crc
\phi  &=& {b^3 \over s}\,\ln\left({ s+\sqrt{s^2+r^2} \over 2s }\right) \\
\phir &=& {b^3 \over s r}\,
              \left[1-{s \over \sqrt{s^2+r^2}}\right] \\
\ell
\phi  &=& {b^3 q \over s}\,\ln\left[(\psi+s)^2+(1-q^2)x^2\right]^{1/2}
              - {b^3 q \over s}\,\ln\left[(1+q)s\right] \\
\phix &=& {b^3 q x \over s \psi}\,{ \psi + q^2 s \over
              (\psi+s)^2 + (1-q^2) x^2 } \\
\phiy &=& {b^3 q y \over s \psi}\,{ \psi + s \over
              (\psi+s)^2 + (1-q^2) x^2 }
\end{eqnarray}
where $\psi^2 = q^2(s^2+x^2)+y^2$. The elliptical solutions are
given by Keeton \& Kochanek (1998).

{\bf Pseudo-Jaffe ellipsoid:}
A standard Jaffe (1983) model has a 3-d density distribution $\rho
\propto r^{-2} (r+a)^{-2}$ where $a$ is the break radius. For
lensing it is useful to modify this model and write $\rho \propto
(r^2+s^2)^{-1} (r^2+a^2)^{-1}$, where $a$ is again the break radius
and we have added a core radius $s < a$. The projected surface
density of the elliptical model has the form
\begin{equation} \label{eq:pjaffe}
  \kappa(\xi) = {b \over 2}\,\left[ {1 \over \sqrt{s^2+\xi^2}} -
    {1 \over \sqrt{a^2+\xi^2}} \right] ,
\end{equation}
which is constant inside $s$, falls as $R^{-1}$ between $s$ and
$a$, and falls as $R^{-3}$ outside $a$; the total mass is $M = \pi
\Sigcr q b (a-s)$. \refEq{pjaffe} defines the pseudo-Jaffe
ellipsoid. In the limit $a \to \infty$ it reduces to the isothermal
ellipsoid ($\al=1$). In the limit $a \to s$ it reduces to the
$\al=-1$ ellipsoid, although the limit must be taken in $\rho$
rather than in $\kappa$ (i.e., the limit must be taken before
the projection integral is evaluated). The pseudo-Jaffe model
is equivalent to a combination of two softened isothermal
ellipsoids, so its lensing properties can be computed with
appropriate combinations of
eqs.~(\ref{eq:isophi})--(\ref{eq:isophiy}).

{\bf King model:}
The King model can be approximated as a combination of two softened
isothermal models (see Young et al.\ 1980; Barkana et al.\ 1999),
\begin{equation}
  \kappa(\xi) = { 2.12\,b \over \sqrt{0.75r_s^2+\xi^2} } -
    { 1.75\,b \over \sqrt{2.99r_s^2+\xi^2} }\ .
\end{equation}
It has a single scale radius $r_s$. This approximation is
convenient because it is written as the difference of two softened
isothermal ellipsoids, so its lensing properties can be computed
with appropriate combinations of
eqs.~(\ref{eq:isophi})--(\ref{eq:isophiy}).

{\bf De Vaucouleurs model:}
This is the prototypical constant mass-to-light ratio lens model
(de Vaucouleurs 1948), with surface mass density
\begin{equation}
  \kappa(\xi) = \kappa_0\,\exp\left[-k(\xi/R_e)^{1/4}\right] ,
\end{equation}
where $k=7.66925001$ and $R_e$ is the major-axis effective (or
half-mass) radius. The circular deflection is (Maoz \& Rix 1993)
\begin{eqnarray}
\phir &=& \kappa_0\,{40320 \over k^8}\,{R_e^2 \over r}\,\Biggl[1- \\
       && \quad e^{-\zeta}
          \left(1+\zeta
          \left(1+{\zeta \over 2}
          \left(1+{\zeta \over 3}
          \left(1+{\zeta \over 4}
          \left(1+{\zeta \over 5}
          \left(1+{\zeta \over 6}
          \left(1+{\zeta \over 7}
          \right)\right)\right)\right)\right)\right)\right) \Biggr] ,
         \nonumber
\end{eqnarray}
where $\zeta = k\,(r/R_e)^{1/4}$. The elliptical model can be
computed numerically with eqs.~(\ref{eq:phi1d})--(\ref{eq:phixy1d}).

{\bf Hernquist model:}
The Hernquist (1990) model is a 3-d density distribution with a
projected distribution that mimics the luminosity distribution of
early-type galaxies. It has the form
\begin{equation}
  \rho = {\rho_s \over (r/r_s)(1+r/r_s)^3}\ ,
\end{equation}
where $r_s$ is a scale length and $\rho_s$ is a characteristic
density. The projected surface mass density has the form
\begin{equation} \label{eq:hernk}
  \kappa(r) = {\kappa_s \over (x^2-1)^2}\,\left[ -3
    + (2+x^2) {\cal F}(x) \right] ,
\end{equation}
where $x = r/r_s$, $\kappa_s = \rho_s\,r_s / \Sigcr$, and
${\cal F}(x)$ is the function
\begin{equation}
{\cal F}(x) = \cases{
  {1 \over \sqrt{x^2-1}}\,\mbox{tan}^{-1} \sqrt{ x^2-1 } & $(x>1)$ \cr
  {1 \over \sqrt{1-x^2}}\,\mbox{tanh}^{-1}\sqrt{ 1-x^2 } & $(x<1)$ \cr
  1                                                      & $(x=1)$ \cr
}
\end{equation}
A useful technical result is the derivative of this function,
\begin{equation}
  {\cal F}'(x) = { 1 - x^2 {\cal F}(x) \over x(x^2-1) }\ .
\end{equation}
The circular potential and deflection are
\begin{eqnarray}
  \phi  &=& \kappa_s\,r_s^2\, \left[ \ln{x^2 \over 4} + 2 {\cal F}(x) \right] ,
  \\
  \phir &=& 2\,\kappa_s\,r_s\,{ x [ 1 - {\cal F}(x) ] \over x^2-1 }\ ,
\end{eqnarray}
where again $x = r/r_s$. The elliptical model $\kappa(\xi)$ can be
computed numerically with eqs.~(\ref{eq:phi1d})--(\ref{eq:phixy1d}).

{\bf NFW model:}
Cosmological $N$-body simulations (e.g., Navarro et al.\ 1996,
1997) suggest that dark matter halos can be described by a
``universal'' density profile with the form
\begin{equation}
  \rho = {\rho_s \over (r/r_s)(1+r/r_s)^2}\ .
\end{equation}
For the spherical NFW model, the projected surface mass density,
potential, and deflection are (Bartelmann 1996; Meneghetti, Bartelmann
\& Moscardini 2001, 2002; Golse \& Kneib 2001)
\begin{eqnarray} \label{eq:nfwk}
  \kappa(r) &=& 2\,\kappa_s\,{1 - {\cal F}(x) \over x^2 - 1}\,,
    \label{eq:nfwkap} \\
  \phi &=& 2\,\kappa_s\,r_s^2\, \left[ \ln^2{x \over 2} -
    \mbox{arctanh}^2 \sqrt{1-x^2} \right] , \\
  \phir &=& 4\,\kappa_s\,r_s\, { \ln(x/2) + {\cal F}(x) \over x }\ ,
    \label{eq:nfwpot}
\end{eqnarray}
where $x = r/r_s$, $\kappa_s = \rho_s\,r_s / \Sigcr$, and the
function ${\cal F}(x)$ is the same as in the Hernquist model.
The elliptical model $\kappa(\xi)$ can be computed numerically
with eqs.~(\ref{eq:phi1d})--(\ref{eq:phixy1d}).

{\bf Cuspy NFW model:}
Moore et al.\ (1998, 1999) have suggested that the inner cusp of
the NFW profile is too shallow, so Jing \& Suto (2000), Keeton \&
Madau (2001), and Wyithe et al.\ (2001) have studied a generalized
NFW-type profile of the form
\begin{equation} \label{eq:nfwcusp}
  \rho = {\rho_s \over (r/r_s)^\gamma (1+r/r_s)^{3-\gamma}}\ ,
\end{equation}
so the central cusp has $\rho \propto r^{-\gamma}$. The projected
surface density cannot be computed analytically even for a spherical
halo. For a spherical model, \refeqs{kap1}{def1} allow the surface
density and deflection to be written as
\begin{eqnarray}
  \kappa(r) &=& 2\,\kappa_s\,x^{1-\gamma}\,
    \left[ (1+x)^{\gamma-3} + (3-\gamma) \int_0^1 dy\,
    (y+x)^{\gamma-4} \left(1-\sqrt{1-y^2}\right) \right] ,
    \label{eq:nfwcuspk} \\
  \phir &=& 4\,\kappa_s\,r_s\,x^{2-\gamma}\, \times \\
  && \Biggl\{\
    {1 \over 3-\gamma}\ {}_2 F_1[ 3-\gamma , 3-\gamma ; 4-\gamma ;
    -x ] + \int_{0}^{1} dy\, (y+x)^{\gamma-3}\,
    { 1-\sqrt{1-y^2} \over y }\ \Biggr\} , \nonumber
\end{eqnarray}
where $x = r/r_s$, $\kappa_s = \rho_s\,r_s/\Sigcr$, and ${}_2 F_1$
is the hypergeometric function.  The integrals can be evaluated for
several values of $\gamma$ to obtain simple analytic expressions.
For $\gamma=1$ the model reduces to the standard NFW model (see
eqs.~\ref{eq:nfwkap} and \ref{eq:nfwpot}).  Two other analytic
cases are as follows:
\begin{eqnarray}
\gamma=2:\quad
  \kappa(r) &=& \kappa_s\,\left[ {\pi \over x} - 2 {\cal F}(x) \right] , \\
  \phir &=& 4\,\kappa_s\,r_s\,\left[ {\pi \over 2}
    + {1 \over x}\,\ln\left({x \over 2}\right) + {1-x^2 \over x}\,{\cal F}(x)
    \right] , \\
\gamma=0:\quad
  \kappa(r) &=& \kappa_s\, { 1 + x^2 [2-3{\cal F}(x)] \over (x^2-1)^2 }\ , \\
  \phir &=& {2\,\kappa_s\,r_s \over x}\,\left[ 2 \ln\left({x \over 2}\right)
    + {x^2 + (2-3x^2){\cal F}(x) \over 1-x^2} \right] .
\end{eqnarray}

Elliptical NFW-cusp models $\kappa(\xi)$ can be computed numerically with
eqs.~(\ref{eq:phi1d})--(\ref{eq:phixy1d}).  In general they are slow to
compute because they require double integrals.

{\bf Cuspy halo models:}
To obtain a general cuspy model that is more amenable to lensing,
Mu\~noz et al.\ (2001) introduce a model with a profile of the
form
\begin{equation}
  \rho = {\rho_s \over (r/r_s)^\gamma [1+(r/r_s)^2]^{(n-\gamma)/2}}\ ,
\end{equation}
where again $r_s$ is a scale length, and $\gamma$ and $n$ are the
logarithmic slopes at small and large radii, respectively.  This
model is a subset of the models whose physical properties were
studied by Zhao (1996).  The central cusp must have $\gamma < 3$
for the mass to be finite.  For $(\gamma,n) = (1,4)$ this is a
pseudo-Hernquist model, for $(1,3)$ it is a pseudo-NFW model, and
for $(2,4)$ it is a singular pseudo-Jaffe model.  Compared with
\refeq{nfwcusp}, replacing $(1+r/r_s)$ with $\sqrt{1+(r/r_s)^2}$
does not greatly change the profile shape but does make it possible
to solve the spherical model analytically (Mu\~noz et al.\ 2001),
\begin{eqnarray}
  \kappa(r) &=& \kappa_s\,B\left( {n-1 \over 2} , {1 \over 2} \right)\,
    \left(1+x^2\right)^{(1-n)/2} {}_2 F_1\left[ {n-1 \over 2} ,
    {\gamma \over 2} ; {n \over 2} ; {1 \over 1+x^2} \right] ,
    \label{eq:cuspk} \\
  n \neq 3: \quad 
  \phir &=& 2\, {\kappa_s r_s \over x} \Biggl\{
    B\left( {n-3 \over 2} , {3-\gamma \over 2} \right)
    \label{eq:cuspphir} \\
    && \qquad - B\left( {n-3 \over 2} , {3 \over 2} \right)
      \left(1+x^2\right)^{(3-n)/2}
      {}_2 F_1 \left[ {n-3 \over 2} , {\gamma \over 2} ;
      {n \over 2} ; {1 \over 1+x^2} \right]
    \Biggr\} , \nonumber
\end{eqnarray}
where $x = r/r_s$, $\kappa_s = \rho_s\,r_s/\Sigcr$, ${}_2 F_1$ is
the hypergeometric function, $B(a,b) = \Gamma(a) \Gamma(b) / \Gamma(a+b)$
is the Euler beta function.  For $n=3$ \refeq{cuspphir} is not valid
(it has singularities of the form 0/0), so it must be replaced with
one of the two following expressions:
\begin{eqnarray}
  n = 3: \quad
  \phir &=& 2 \kappa_s r_s \Biggl\{ {1 \over x}\,\ln(1+x^2)
    - G\left[ {\gamma \over 2} , {\gamma-1 \over 2} ; {x^2 \over 1+x^2} \right]
    \label{eq:cusp3a} \\
  && \qquad - x^{2-\gamma} (1+x^2)^{(\gamma-3)/2}\,
      B\left( {\gamma-3 \over 2} , {3 \over 2} \right)
      {}_2 F_1 \left[ {3 \over 2} , {3-\gamma \over 2} ;
      {5-\gamma \over 2} ; {x^2 \over 1+x^2} \right] \Biggr\}
    \nonumber \\
  &=& 2\, {\kappa_s r_s \over x} \left\{ \ln(1+x^2)
    - G\left[ {\gamma \over 2} , {3 \over 2} ; {1 \over 1+x^2} \right]
    + \Psi\left({3 \over 2}\right) - \Psi\left({3-\gamma \over 2}\right)
    \right\} \label{eq:cusp3b}
\end{eqnarray}
where $\Psi(x)$ is the digamma function (see eq.~\ref{eq:alphapot}),
and $G(b,c;z)$ is the function
\begin{eqnarray}
  G(b,c;z) &\equiv& \lim_{a \to 0} { {}_2 F_1 [a,b,c;z] - 1 \over a } \\
  &=& {b \over c}\,z + {b(b+1) \over c(c+1)}\,{z^2 \over 2} +
    {b(b+1)(b+2) \over c(c+1)(c+2)}\,{z^3 \over 3} + \cdots \\
  &=& \sum_{j=1}^{\infty} { b(b+1)\cdots(b+j-1) \over c(c+1)\cdots(c+j-1) }\,
    {z^j \over j}
\end{eqnarray}
Note that \refeqs{cusp3a}{cusp3b} are equivalent and can be transformed
into each other using identities for the hypergeometric function (see
Gradshteyn \& Rhyzik 1994, \S 9.131).  It is convenient to use \refeq{cusp3a}
for $x<1$ and \refeq{cusp3b} for $x>1$ in order to make the series for the
$G$ function converge rapidly.

Elliptical cusp models $\kappa(\xi)$ can be computed numerically with
eqs.~(\ref{eq:phi1d})--(\ref{eq:phixy1d}), or with a Fourier series
solution (Chae 2002).

{\bf Nuker law:}
Many early-type galaxies have surface brightness profiles that can be
modeled as a Nuker law (e.g., Lauer et al.\ 1995; Byun et al.\ 1996),
\begin{equation}
  I(r) = 2^{(\beta-\gamma)/\alpha}\, I_b\, \left({r \over r_b}\right)^{-\gamma}
    \left[1+\left({r \over r_b}\right)^{\alpha}\right]^{(\gamma-\beta)/\alpha} ,
    \label{eq:nukerk}
\end{equation}
where $\gamma$ and $\beta$ are inner and outer power law indices,
respectively, $r_b$ is the radius where the break in the power law
occurs, $\alpha$ gives the sharpness of the break, and $I_b$ is the
surface brightness at the break radius.  If the luminosity distribution
has circular symmetry and the mass-to-light ratio is $\Upsilon$, the
lensing deflection is (Keeton 2002)
\begin{equation}
  \phir = {2^{1+(\beta-\gamma)/\alpha} \over 2-\gamma}\
    \kappa_b\, r_b\, \left({r \over r_b}\right)^{1-\gamma}\
    {}_2 F_1\left[ {2-\gamma \over \alpha} , {\beta-\gamma \over \alpha} ;
    1 + {2-\gamma \over \alpha} ; -\left({r \over r_b}\right)^{\alpha} \right] ,
\end{equation}
where $\kappa_b = \Upsilon I_b/\Sigcr$ is the surface mass density
at the break radius in units of the critical density for lensing,
and ${}_2 F_1$ is the hypergeometric function.  Ellipsoidal models
can be computed numerically with
eqs.~(\ref{eq:phi1d})--(\ref{eq:phixy1d}).

{\bf Exponential disk:}
The projected surface density is
\begin{equation}
  \kappa(\xi) = q^{-1}\,\kappa_0\,\exp\left[-\xi/R_d\right],
\end{equation}
which represents a thin, circular disk with intrinsic central
density $\kappa_0$ and scale length $R_d$, seen in projection with
axis ratio $q=|\cos i|$ where $i$ is the inclination angle (such
that $i=0\arcdeg$ is face-on and $i=90\arcdeg$ is edge-on). The
circular deflection is
\begin{equation}
  \phir = 2\,\kappa_0\,{R_d^2 \over r}\,\left[1-
    \left(1+{r \over R_d}\right)\,e^{-r/R_d}\right] .
\end{equation}
The elliptical model $\kappa(\xi)$ can be computed with
eqs.~(\ref{eq:phi1d})--(\ref{eq:phixy1d}), or it can be
approximated with one or more Kuzmin disks (see Keeton \& Kochanek
1998).

{\bf Kuzmin disk:}
The $\al = -1$ ellipsoid can be re-interpreted as the projection of
a thin disk, in which case it corresponds to a Kuzmin (1956) or
Toomre (1962) Model I disk; see Keeton \& Kochanek (1998). Its
projected surface density is
\begin{equation}
  \kappa(\xi) = q^{-1}\,\kappa_0\,r_s^3\,\left(r_s^2+\xi^2\right)^{-3/2} ,
\end{equation}
where $\kappa_0$ is the intrinsic central surface density of the
disk, and $q=|\cos i|$ is again the projected axis ratio of the
inclined disk. The only difference between the Kuzmin disk and the
$\al = -1$ ellipsoid is the normalization.

{\bf External perturbations:}
Objects near the main lens galaxy or along the line of sight often
perturb the lensing potential. If the perturbation is weak it may
be sufficient to expand the perturbing potential as a Taylor series
and keep only a few terms. In a coordinate system centered on the
lens galaxy, the expansion to 3rd order can be written as (see
Kochanek 1991; Bernstein \& Fischer 1999)
\begin{equation}
  \phi \approx \phi_0 + \bb \cdot \xx
    + {r^2 \over 2}\,\Bigl[ \kappa - \gamma\,\cos2(\th-\th_\gamma) \Bigr]
    + {r^3 \over 3}\,\Bigl[ \delta\,\cos(\th-\th_\delta) -
      \varepsilon\,\cos3(\th-\th_\varepsilon) \Bigr] + \ldots
\end{equation}
The 0th order term $\phi_0$ represents an unobservable zero point
of the potential and can be dropped. The 1st order term $\bb \cdot
\xx$ represents an unobservable uniform deflection and can also be
dropped. The 2nd order term $\kappa$ represents the convergence
from the perturbing mass and is equivalent to a uniform mass sheet
with density $\Sigma/\Sigcr = \kappa$. The only observable effect
of this term is to rescale the time delay(s) by $1-\kappa$, which
leads to the ``mass sheet degeneracy'' (e.g., Falco, Gorenstein \&
Shapiro 1985); hence this term is often omitted from lens models
and introduced {\it a posteriori\/} using independent mass constraints
(see, e.g., Bernstein \& Fischer 1999). The 2nd order term $\gamma$
represents an external tidal shear with strength $\gamma$ and direction
$\th_\gamma$. The 3rd order term $\delta$ arises from the gradient of
the surface density $\kappa(\xx)$ of the perturber; it has an amplitude
$\delta = (3/4) |\del\kappa|$ and a direction equal to the direction
of $\del\kappa$. The 3rd order term $\varepsilon$ arises from the
$m=3$ multipole moment of the perturbing mass. The constant
coefficients $(\kappa,\gamma,\delta,\varepsilon)$ are all
evaluated at the position of the lens galaxy, and the corresponding
direction angles are written here as theory angles measured
counter-clockwise from the $x$-axis.

\acknowledgements
Acknowledgements: I would like to thank Chris Kochanek, Joanne Cohn,
Jose Mu\~noz, David Rusin, Brian McLeod, and Joseph Leh\'ar for
many helpful discussions about the models and about the \gravlens\
software.  Support for this work has been provided by ONR-NDSEG
grant N00014-93-I-0774, NSF grant AST-9407122, NASA ATP grant
NAG5-4062, Steward Observatory, and Hubble Fellowship grant
HST-HF-01141.01-A from the Space Telescope Science Institute, which
is operated by the Association of Universities for Research in
Astronomy, Inc., under NASA contract NAS5-26555.



\clearpage

\begin{deluxetable}{lcll}
\tablewidth{0pt}
\tablecaption{Mass Models for Lensing}
\tablehead{
  \colhead{Model} & \colhead{$N_r$}
& \colhead{Density $\rho(r)$}
& \colhead{Surface Density $\kappa(r)$}
}
\startdata
Point mass & 0
 & $\delta(\xx)$
 & $\delta(\xx)$ \\
Power law or $\al$-models & 2
 & $\left(s^2+r^2\right)^{(\al-3)/2}$
 & $\left(s^2+r^2\right)^{(\al-2)/2}$ \\
Isothermal ($\al=1$) & 1
 & $\left(s^2+r^2\right)^{-1}$
 & $\left(s^2+r^2\right)^{-1/2}$ \\
$\al=-1$ & 1
 & $\left(s^2+r^2\right)^{-2}$
 & $\left(s^2+r^2\right)^{-3/2}$ \\
Pseudo-Jaffe & 2
 & $\left(s^2+r^2\right)^{-1} \left(a^2+r^2\right)^{-1}$
 & $\left(s^2+r^2\right)^{-1/2} - \left(a^2+r^2\right)^{-1/2}$ \\
King (approximate) & 1
 & \nodata
 & $2.12 \left(0.75 r_s^2+r^2\right)^{-1/2}$ \\
&&& $\quad - 1.75 \left(2.99 r_s^2+r^2\right)^{-1/2}$ \\
de Vaucouleurs & 1
 & \nodata
 & $\exp\left[-7.67 (r/R_e)^{1/4}\right]$ \\
Hernquist & 1
 & $r^{-1} \left(r_s+r\right)^{-3}$
 & see \refeq{hernk} \\
NFW & 1
 & $r^{-1} \left(r_s+r\right)^{-2}$
 & see \refeq{nfwk} \\
Cuspy NFW & 2
 & $r^{-\gamma} \left(r_s+r\right)^{\gamma-3}$
 & see \refeq{nfwcuspk} \\
Cusp & 3
 & $r^{-\gamma} \left(r_s^2+r^2\right)^{(\gamma-n)/2}$
 & see \refeq{cuspk} \\
Nuker & 4
 & \nodata
 & see \refeq{nukerk} \\
Exponential disk & 1
 & \nodata
 & $\exp[-r/R_d]$ \\
Kuzmin disk & 1
 & \nodata
 & $\left(r_s^2+r^2\right)^{-3/2}$ \\
\enddata
\tablecomments{
Density profiles for lensing mass models; see \S 5 for detailed
definitions, including normalizations.  Three-dimensional density
profiles are not given for the King, de Vaucouleurs, Nuker,
exponential disk, and Kuzmin disk models because these models are
defined by their surface densities.  Column 2 indicates the number
of parameters associated with the radial profile alone; each model
would also have parameters for the position and the mass scale,
and elliptical models would have parameters for the ellipticity
and orientation.  The profiles are given for spherical models;
elliptical models are defined by $\kappa(\xi)$ where $\xi$ is an
ellipse coordinate (see eq.~\ref{eq:xi}).
}\label{tab:mods}
\end{deluxetable}


\begin{references}

\reference{}
Barkana, R. 1998, \apj, 502, 531

\reference{}
Barkana, R., Leh\'ar, J., Falco, E. E., Grogin, N. A., Keeton, C. R.,
\& Shapiro, I. I. 1999, \apj, 523, 54

\reference{}
Bartelmann, M., \& Steinmetz, M. 1995, \aap, 297, 1

\reference{}
Bartelmann, M. 1996, \aap, 313, 697

\reference{}
Bernstein, G., \& Fischer, P. 1999, \aj, 118, 14

\reference{}
Binney, J., \& Tremaine, S. 1987, Galactic Dynamics (Princeton:
Princeton Univ. Press)

\reference{}
Blandford, R. D., \& Kochanek, C. S. 1987, \apj, 321, 658

\reference{}
Byun, Y.-I., et al. 1996, \aj, 111, 1889

\reference{}
Chae, K.-H. 2002, \apj, 568, 500

\reference{}
Cohn, J. D., Kochanek, C. S., McLeod, B. A., \& Keeton, C. R. 2001,
\apj, 554, 1216

\reference{}
de Vaucouleurs, G. 1948, Ann d'Ap, 11 247

\reference{}
Evans, N. W., \& Hunter, C. 2002, preprint (astro-ph/0204206)

\reference{}
Fabbiano, G. 1989, \araa, 27, 87

\reference{}
Faber, S. M., Tremaine, S., Ajhar, E. A., Byun, Y.-I., Dressler, A.,
Gebhardt, K., Grillmair, C., Kormendy, J., Lauer, T. R., \& Richstone, D.
1997, \aj, 114, 1771

\reference{}
Falco, E. E., Gorenstein, M. V., \& Shapiro, I. I. 1985, \apj, 289, L1

\reference{}
Golse, G., \& Kneib, J.-P. 2001, preprint (astro-ph/0112138)

\reference{}
Gould, A., \& Loeb, A. 1992, \apj, 396, 104

\reference{}
Gradshteyn, I. S., \& Ryzhik, I. M. 1994, Table of Integrals, Series,
and Products, Fifth Edition, ed. A. Jeffrey (San Diego: Academic Press)

\reference{}
Hernquist, L. 1990, \apj, 356, 359

\reference{}
Jaffe, W. 1983, \mnras, 202, 995

\reference{}
Jing, Y. P., \& Suto, Y. 2000, \apj, 529, L69

\reference{}
Kassiola, A., \& Kovner, I. 1993, \apj, 417, 459

\reference{}
Keeton, C. R., Kochanek, C. S., \& Seljak, U. 1997, \apj, 482, 604

\reference{}
Keeton, C. R., \& Kochanek, C. S. 1997, \apj, 487, 42

\reference{}
Keeton, C. R., \& Kochanek, C. S. 1998, \apj, 495, 157

\reference{}
Keeton, C. R. 1998, PhD. thesis, Harvard University

\reference{}
Keeton, C. R., Kochanek, C. S., \& Falco, E. E. 1998, \apj, 509, 561

\reference{}
Keeton, C. R., Falco, E. E., Impey, C. D., Kochanek, C. S., Leh\'ar, J.,
McLeod, B. A., Rix, H.-W., Mu\~noz, J. A., \& Peng, C. Y. 2000,
\apj, 542, 74

\reference{} 
Keeton, C. R. 2001, \apj, preprint (astro-ph/0102340)

\reference{} 
Keeton, C. R., \& Madau, P. 2001, \apj, 549, L25

\reference{} 
Keeton, C. R. 2002, \apj, submitted

\reference{} 
Kochanek, C. S. 1991, \apj, 382, 58

\reference{} 
Kochanek, C. S. 1993, \apj, 419, 12

\reference{} 
Kochanek, C. S. 1995, \apj, 445, 559

\reference{} 
Kochanek, C. S. 1996, \apj, 466, 638

\reference{} 
Kochanek, C. S., \& White, M. 2001, \apj, 559, 531

\reference{}
Koopmans, L. V. E., \& Fassnacht, C. D. 1999, \apj, 527, 513

\reference{}
Koopmans, L. V. E., \& Treu, T. 2002, preprint (astro-ph/0205281)

\reference{}
Kormann, R., Schneider, P., \& Bartelmann, M. 1994, \aap, 284, 285

\reference{}
Kuzmin, G. 1956, AZh, 33, 27

\reference{}
Lauer, T. R., et al. 1995, \aj, 110, 2622

\reference{}
Maller, A. H., Flores, R. A., \& Primack, J. R. 1997, \apj, 486, 681

\reference{}
Mao, S., \& Paczy\'nski, B. 1991, \apj, 374, L37

\reference{}
Maoz, D., \& Rix, H.-W. 1993, \apj, 416, 425

\reference{}
Meneghetti, M., Bartelmann, M., \& Moscardini, L. 2001, preprint
(astro-ph/0109250)

\reference{}
Meneghetti, M., Bartelmann, M., \& Moscardini, L. 2002, preprint
(astro-ph/0201501)

\reference{}
Moore, B., Governato, F., Quinn, T., Stadel, J., \& Lake, G. 1998,
\apj, 499, L5

\reference{}
Moore, B., Quinn, T., Governato, F., Stadel, J., \& Lake, G. 1999,
\mnras, 310, 1147

\reference{}
Mu\~noz, J. A., Kochanek, C. S., \& Keeton, C. R. 2001, \apj, 558, 657

\reference{}
Navarro, J. F., Frenk, C. S., \& White, S. D. M. 1996, \apj, 462, 563

\reference{}
Navarro, J. F., Frenk, C. S., \& White, S. D. M. 1997, \apj, 490, 493

\reference{}
Porciani, C., \& Madau, P. 2000, \apj, 532, 679

\reference{}
Press, W. H., Teukolsky, S. A., Vetterling, W. T., \& Flannery, B. P.
1992, Numerical Recipes in C: The Art of Scientific Computing, Second
Edition (New York: Cambridge Univ. Press)

\reference{}
Rix, H.-W., de Zeeuw, P. T., Carollo, C. M., Cretton, N., \&
van der Marel, R. P. 1997, \apj, 488, 702

\reference{}
Rubin, V. C., Ford, W. K., \& Thonnard, N. 1978, \apj, 225, L107

\reference{}
Rubin, V. C., Ford, W. K., \& Thonnard, N. 1980, \apj, 238, 471

\reference{}
Rusin, D., \& Ma, C.-P. 2001, \apj, 549, L33

\reference{}
Saha, P., \& Williams, L. L. R. 1997, \mnras, 292, 148

\reference{}
Schechter, P. L., et al. 1997, \apj, 475, L85

\reference{}
Schneider, P., \& Weiss, A. 1991, \aap, 247, 269

\reference{}
Schneider, P., Ehlers, J., \& Falco, E. E. 1992, Gravitational Lenses
(New York: Springer)

\reference{}
Schramm, T. 1990, \aap, 231, 19

\reference{}
Toomre, A. 1962, \apj, 138, 385

\reference{}
Treu, T., \& Koopmans, L. V. E. 2002, preprint (astro-ph/0202342)

\reference{}
Tyson, J. A., Kochanski, G. P., \& Dell'Antonio, I. P. 1998, \apj,
498, L107

\reference{}
Wallington, S., \& Narayan, R. 1993, \apj, 403, 517

\reference{}
Williams, L. L. R., \& Saha, P. 2000, \aj, 119, 439

\reference{}
Witt, H. J., \& Mao, S. 1997, \mnras, 291, 211

\reference{}
Witt, H. J., \& Mao, S. 2000, \mnras, 311, 689

\reference{}
Wyithe, J. S. B., Turner, E. L., \& Spergel, D. N. 2001, \apj, 555, 504

\reference{}
Young, P., Gunn, J. E., Kristian, J., Oke, J. B., \& Westphal, J. A.
1980, \apj, 241, 507

\reference{}
Young, P., Gunn, J. E., Kristian, J., Oke, J. B., \& Westphal, J. A.
1981, \apj, 244, 736

\reference{}
Zhao, H.-S. 1996, \mnras, 278, 488

\end{references}
\end{document}